\documentclass[final,1p,times]{elsarticle} % do not change
\usepackage{graphicx} % do not change
\usepackage{amssymb} % do not change
\usepackage{amsthm} % do not change
\usepackage{lineno} % do not change

\journal{Nuclear Physics A} % do not change
\begin{document} % do not change

\begin{frontmatter} % do not change

%% QM09Author: please enter your  
%% Title, author and address info here; please do not use footnotes

% Your Title - please insert
\title{Langevin + Hydrodynamics Approach to Heavy Quark Propagation and Correlation in QGP}

% Principle author, and co-authors - please insert
\author{Yukinao Akamatsu, Tetsuo Hatsuda, Tetsufumi Hirano}

% Address - please insert
\address{Department of Physics, The University of Tokyo, 113-0033, Japan}

\begin{abstract} % do not change
%% Text of abstract goes here - please insert
We develop a relativistic Langevin dynamics under the background of strongly interacting quark-gluon fluid described by the (3+1)-dimensional hydrodynamics.
The drag force acting on charm and bottom quarks is parametrized according to the formula obtained from the anti-de-Sitter space/conformal field theory (AdS/CFT) correspondence.
In this setup, we calculate the nuclear modification factor $R_{\rm{AA}}$ for the single-electrons from the charm and bottom quarks to extract the magnitude of the drag force from the PHENIX and STAR data.
The $R_{\rm{AA}}$ for single-electrons with $p_{T}\geq 3$ GeV indicates that the drag force is close to the AdS/CFT prediction.  
Effects of the drag force on the elliptic flow of single-electrons are also discussed.
Moreover, we predict the electron-muon correlation which is closely related to the heavy-quark pair correlation in hot matter.
\end{abstract} % do not change

\end{frontmatter} % do not change

%% QM09: we keep linenumbers at least for initial version
%\linenumbers % do not change

%% start of main text - please insert. 

\section{Introduction}\label{intro}

The properties of quark-gluon plasma (QGP) are actively investigated by relativistic heavy ion collision experiments at Relativistic Heavy Ion Collider (RHIC) and will be studied at the Large Hadron Collider (LHC) in the future.
The space-time evolution of the created hot medium is described nicely by the relativistic hydrodynamic model in which vanishing viscosity is assumed.
This fact implies the strongly interacting nature of the QGP even above the deconfinement transition temperature $T_c \sim$ 170 MeV.
In heavy ion collisions, majority of the particles, namely light soft particles such as up, down and strange quarks and gluons, constitute the hot medium whose space-time evolution is described by the hydrodynamic model.
However, heavy quarks, namely charm and bottom quarks, do not join the medium but act as impurities in the medium because they have much heavier masses ($m_c \simeq  1.5$ GeV, $m_b \simeq  4.8$ GeV) than typical temperature available at RHIC ($T \sim 300$ MeV). 
Not only heavy quarks but also jets and $J/\Psi$s act as impurities in the medium.

In this report, we focus on heavy quarks in the QGP, especially on their diffusive dynamics in the medium and on its influence on the spectra of their observables in experiments, namely single-leptons decayed from heavy mesons.
We perform simulations of heavy quark diffusion \cite{akamatsu} in the background of expanding hot medium and of the subsequent heavy quark hadronization followed by semileptonic decay and make a quantitative comparison with experimental data \cite{experiment}.

\section{Langevin + Hydrodynamics Model}\label{model}
Since masses of heavy quarks are much larger than the typical temperature of surrounding medium while kinematics of heavy quarks can be relativistic in actual experiments, the heavy quark dynamics may be described by the relativistic Langevin equation.
The basic equations in the rest frame of matter are
\begin{eqnarray}
\label{eq:langevin1}
&&\Delta \vec x = \frac{\vec p}{E}\Delta t, \ \ 
\Delta\vec p =-\Gamma(p)\vec p\Delta t +\vec \xi(t),\\
\label{eq:langevin2}
&&\langle\xi _{i}(t)\xi _{j}(t')\rangle =D_{ij}(p)\delta _{tt'}\Delta t,
\end{eqnarray}
where we use the specific parametrization, $\Gamma(p)=\gamma T^2/M$ and $D_{ij}(p)=2\gamma T^3 (E+T)\delta _{ij}/M$, motivated by the drag force calculated on the basis of the anti-de-Sitter space/conformal field theory (AdS/CFT) correspondence \cite{ads_drag}.
According to an attempt to translate the drag parameter $\gamma$ of the supersymmetric Yang-Mills plasma into the one in strongly interacting QGP, the drag parameter is estimated to be $\gamma= 2.1\pm 0.5$ \cite{translation}.

The combination of the relativistic Langevin dynamics of heavy quarks and hydrodynamic expansion of hot medium is implemented by the following procedures.
(i) We first specify the initial condition of heavy quarks.
The initial distribution is given by multiplying the nucleon-nucleon collision by the number of binary collisions, namely the Glauber model.
(ii) Heavy quarks diffuse in the expanding medium.
The local flow vector and temperature are given by the hydrodynamic model.
(iii) Heavy quarks hadronize to heavy mesons, namely $D$ or $B$ mesons, via independent fragmentation when confinement phase transition takes place.
(iv) Heavy mesons decay to single-electrons via semileptonic decay.

Some comments are in order here:
(a) The initial heavy quark momentum distribution is computed by PYTHIA event generator, which is based on leading order perturbative QCD and is found to be reliable only at high transverse momentum $p_{T}$ region.
For this reason, we limit ourselves to the comparison of the single-electron results from our simulation with those from experiment only in their spectral shape at $p_T\geq$ 3.0 GeV.
(b) Since we assume first order confinement phase transition in the hydrodynamic model, the mixed phase of QGP and hadronic phase causes ambiguity in the determination of the time when heavy quarks hadronize to heavy mesons.
We treat this ambiguity as a systematic error by performing simulations with different freezeout criteria.

\section{$R_{AA}$ for single electrons and electron-muon azimutal correlation}\label{result}
\begin{figure}[h]
\centering
\vspace{0.2cm}
\includegraphics[width=8cm,origin,clip]{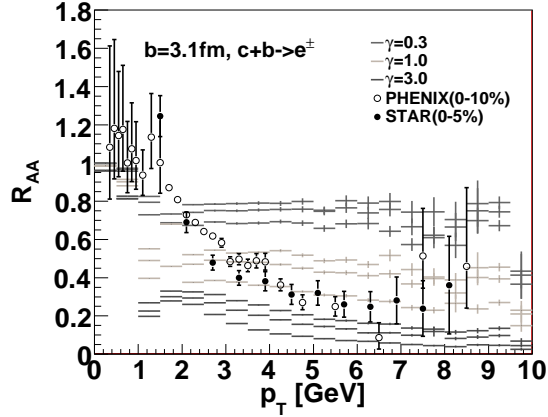}
\caption[raa]
{\footnotesize
Comparison of the nuclear modification factor $R_{\rm{AA}}$ of single-electrons in our hydro + heavy-quark model and the experimental data.
The collision geometry is fixed by the impact parameter 3.1 fm and the single-electrons with mid-rapidity $|\eta_p|\leq$ 0.35 are counted.
Results of our simulation with different drag parameters and with different freezeout conditions are drawn.
}
\label{raa}
\end{figure}

We show in Fig.\ref{raa} the nuclear modification factor $R_{\rm {AA}}$ of single-electrons defined as,
\begin{eqnarray}
R_{\rm AA}(p_T)=\frac{1}{N_{\rm coll}}\frac{dN^e_{\rm A+A}/dp_T}{dN^e_{\rm p+p}/dp_T} \ \ .
\end{eqnarray}
The collision geometry for $R_{\rm{AA}}$ is fixed by impact parameter $b=$ 3.1 fm and the rapidity range considered is $|\eta_p|\leq$ 0.35.
Numerical results of simulation with different drag parameter $\gamma$ and with different choices of freezeout condition are plotted.
By comparing our simulation with experimental data at $p_{T}\geq$ 3.0 GeV, we find that the drag parameter $\gamma=$ 1-3 is favored.
Note that this value is consistent with the prediction from the AdS/CFT correspondence $\gamma =2.1\pm 0.5$.

The degree of thermalization of heavy quarks can be estimated by the ratio of two time scales, the stay time $t_S$ and the relaxation time $\tau_Q$.
With a drag parameter $\gamma =$ 1-3, $t_S$ is 3-4 fm and $\tau_Q$ is 2-7 fm for charm and 7-20 fm for bottom quarks \cite{akamatsu}.
Therefore the charm quarks are partially thermalized while bottom quarks are not thermalized at RHIC.

As for elliptic flow $v_2$ of single-electrons, our simulation and experimental data still suffer from poor statistics at $p_T\geq$ 3.0 GeV and definite conclusions cannot be drawn at the moment.

\begin{figure}[h]
\centering
\vspace{0.2cm}
\includegraphics[width=8cm,origin,clip]{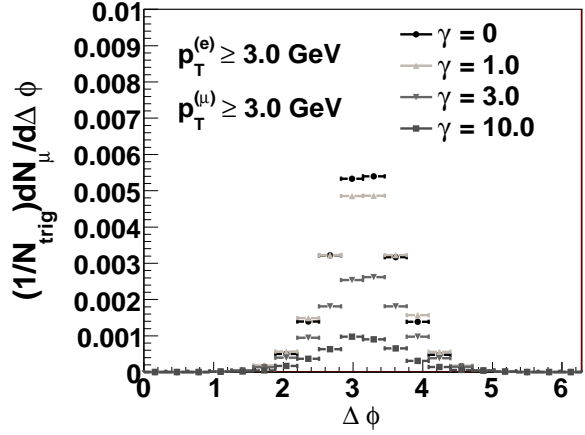}
\caption[corr]
{\footnotesize
Azimuthal correlation of a single-electron and a single-muon decayed from a heavy quark and anti-heavy quark pair.
Collision geometry is fixed by the impact parameter 3.1 fm.
Both trigger electrons and associate muons have transverse momentum $p_{T}\geq 3.0$ GeV and are in mid-pseudorapidity $|\eta|\leq 1.0$. 
}
\label{corr}
\end{figure}

We show in Fig.\ref{corr} our results of electron-muon azimuthal correlation. 
This azimuthal correlation is a clean observable and free of the contamination by di-leptons from vector meson decays.
The trigger particles are single-electrons with $p_T\geq$ 3.0 GeV and the associated particles are single-muons with $p_T\geq$ 3.0 GeV.
The high $p_T$ trigger electron is desirable because it takes over the direction of its parent heavy meson with even higher $p_T$, which also inherits the direction of the heavy quark just before hadronization.
Thus the azimuthal correlation between a heavy quark and an anti-heavy quark pair at their freezeout can be studied by the correlation between a single-electron and a single-muon.
Collision geometry is again fixed by the impact parameter $b$ = 3.1 fm.
We show our results with the pseudorapidity region $|\eta |\leq$ 1.0 in Fig.\ref{corr}.
In this correlation, we can see only one peak in the away side because a single-electron and a single-muon cannot be produced simultaneously in one semileptonic decay.
For the drag parameter $\gamma\geq$ 3, the away side quenching is significant.
Therefore, clean measurement of the azimuthal correlation between a single-electron and a single-muon with high transverse momentum $p_T\geq$ 3.0 GeV can probe the strongly interacting QGP quite well.\\

\section{Summary}\label{summary}

In this report, we showed our studies on the heavy quark diffusion in the dynamical QGP fluid on the basis of the relativistic Langevin equation combined with the relativistic hydrodynamics.
Motivated by the formula from the AdS/CFT correspondence for strongly interacting plasma, we parametrized the drag coefficient by $\Gamma \equiv\gamma T^{2}/M$ with the dimensionless coefficient $\gamma$ as a parameter to be extracted from experiment. 
We obtained single-electron spectra by solving the Langevin equation for heavy quarks on the space-time evolution of background fluid and by simulating their hadronization and subsequent semileptonic decay.

The comparison of $R_{\rm{AA}}$ for $p_T> 3$ GeV suggests that the drag coefficient could be as large as $\gamma=$ 1.0 $-$ 3.0, consistent with the prediction by the AdS/CFT.
We have also found that azimuthal electron-muon correlation with high transverse momentum $p_T>3$ GeV has a significant dependence on the drag parameter and thus can serve as a good probe for the strongly interacting QGP.

%% end of main text

\section*{Acknowledgments} % please insert, comment out or delete if not needed
%This is where one places acknowledgments for funding bodies etc., if needed.
%For the large collaborations, this is listed once and for all, together with 
%the author lists etc. in the proceedings back-material.
Y.~Akamatsu is supported by JSPS fellowships for Young Scientists.
T.~Hatsuda is partially supported by No.~2004, Grant-in-Aid for Scientific Research on Innovative Areas.
T.~Hirano is partially supported by Grant-in-Aid for Scientific Research No.~19740130 and by Sumitomo Foundation No.~080734.

 % do not change 
\end{document}